\documentclass{article}
\usepackage{epsfig}

\date{\today}

\usepackage{amsmath}
\usepackage{amssymb}

\begin{document}
\title{ Holographic Superconductors in 3+1 dimensions away from the probe limit}
\author{{\large Yves Brihaye \footnote{email: yves.brihaye@umons.ac.be} }$^{\ddagger}$ and 
{\large Betti Hartmann \footnote{email: b.hartmann@jacobs-university.de}}$^{\dagger}$
\\ \\
$^{\ddagger}${\small Physique-Math\'ematique, Universite de
Mons-Hainaut, 7000 Mons, Belgium}\\ 
$^{\dagger}${\small School of Engineering and Science, Jacobs University Bremen, 28759 Bremen, Germany}  }

\date{}
\newcommand{\dd}{\mbox{d}}
\newcommand{\tr}{\mbox{tr}}
\newcommand{\la}{\lambda}
\newcommand{\ka}{\kappa}
\newcommand{\f}{\phi}
\newcommand{\vf}{\varphi}
\newcommand{\F}{\Phi}
\newcommand{\al}{\alpha}
\newcommand{\ga}{\gamma}
\newcommand{\de}{\delta}
\newcommand{\si}{\sigma}
\newcommand{\bomega}{\mbox{\boldmath $\omega$}}
\newcommand{\bsi}{\mbox{\boldmath $\sigma$}}
\newcommand{\bchi}{\mbox{\boldmath $\chi$}}
\newcommand{\bal}{\mbox{\boldmath $\alpha$}}
\newcommand{\bpsi}{\mbox{\boldmath $\psi$}}
\newcommand{\brho}{\mbox{\boldmath $\varrho$}}
\newcommand{\beps}{\mbox{\boldmath $\varepsilon$}}
\newcommand{\bxi}{\mbox{\boldmath $\xi$}}
\newcommand{\bbeta}{\mbox{\boldmath $\beta$}}
\newcommand{\ee}{\end{equation}}
\newcommand{\eea}{\end{eqnarray}}
\newcommand{\be}{\begin{equation}}
\newcommand{\bea}{\begin{eqnarray}}

\newcommand{\ii}{\mbox{i}}
\newcommand{\e}{\mbox{e}}
\newcommand{\pa}{\partial}
\newcommand{\Om}{\Omega}
\newcommand{\vep}{\varepsilon}
\newcommand{\bfph}{{\bf \phi}}
\newcommand{\lm}{\lambda}
\def\theequation{\arabic{equation}}
\renewcommand{\thefootnote}{\fnsymbol{footnote}}
\newcommand{\re}[1]{(\ref{#1})}
\newcommand{\R}{{\rm I \hspace{-0.52ex} R}}
\newcommand{\N}{{\sf N\hspace*{-1.0ex}\rule{0.15ex}%
{1.3ex}\hspace*{1.0ex}}}
\newcommand{\Q}{{\sf Q\hspace*{-1.1ex}\rule{0.15ex}%
{1.5ex}\hspace*{1.1ex}}}
\newcommand{\C}{{\sf C\hspace*{-0.9ex}\rule{0.15ex}%
{1.3ex}\hspace*{0.9ex}}}
\newcommand{\eins}{1\hspace{-0.56ex}{\rm I}}
\renewcommand{\thefootnote}{\arabic{footnote}}

\maketitle

\bigskip

\begin{abstract}
We study holographic superconductors in 3+1 dimensions away from the probe limit,
i.e. taking back--reaction of the space-time into account. We consider the case
of pure Einstein - and Gauss--Bonnet gravity, respectively.
Similar to  the probe limit we observe that the critical temperature at which condensation sets in decreases
with increasing Gauss--Bonnet coupling. The decrease is however stronger
when taking back--reaction of the space--time into account. We observe 
that the critical temperature becomes very small, but stays positive
for all values of the Gauss--Bonnet coupling no matter how strong the back--reaction of the
space--time is.

\end{abstract}
\section{Introduction}

The gravity--gauge theory duality \cite{ggdual} has attracted a lot of attention in the past years. The most famous
example is the AdS/CFT correspondence \cite{adscft} which states that a gravity theory in a $d$-dimensional
Anti-de Sitter (AdS) space--time is equivalent to a Conformal Field Theory (CFT) on the $(d-1)$-dimensional boundary of AdS.

Recently, this theory has been used to describe so-called holographic superconductors with the help of black holes in
higher dimensional space--time \cite{gubser,hhh,reviews} and many aspects have
been discussed such as holographic superconductors in Horava-Lifshitz gravity \cite{HL} and
in Born-Infeld electrodynamics \cite{BIholo}, fermions \cite{fermions_holo}, the behaviour
of holographic superconductors in external magnetic fields \cite{exmagnetic} and at 
zero temperature \cite{zeroT}, hydrodynamical aspects of holographic superconductors
\cite{hydro} as well as rotating superconductors \cite{Sonner}.
Holographic superconductors in extended models that allow for a first order phase transition \cite{extended} as well as holographic superconductors in M-Theory \cite{Mtheory} have also been
studied.
Non-abelian (or p-wave) holographic superconductors have been studied in \cite{gubser3,gubser4,aegko,non-abelian,erdmenger2,erdmenger3,basu,herzog_pufu,erdmenger4}. In \cite{erdmenger2,erdmenger3} a string theory realization
of p-wave holographic superconductors in the probe limit has been discussed and the
Meissner effect  has been studied in detail \cite{erdmenger3}.
(For a related analytical study see \cite{basu}.) Sound modes for p-wave superconductors
have been considered in \cite{herzog_pufu}, while fermions in these superconductors have been discussed in \cite{erdmenger4}.
Various other aspects have also been studied \cite{otheraspects}.
 
The general idea behind holographic superconductors comes
from the observation that below a critical temperature electrically charged
black holes become unstable to form scalar hair, i.e. they possess non--vanishing scalar fields on the horizon \cite{gubser}. The reason for this is that
close to the horizon of the black hole the effective mass of the scalar field can become
negative with masses below the Breitenlohner--Freedman bound \cite{bf} such that the scalar
field becomes unstable and possesses a non--vanishing value on and close to the horizon
of the black hole.
The value of the scalar field on the AdS boundary is then associated with the corresponding condensate in the dual theory.

In most cases, holographic superconductors have been studied in the ``probe limit'' neglecting back--reaction of the space--time. This limit corresponds to 
letting the electric charge $e$ tend to infinity or equivalently  Newton's constant $G$ tend to zero.
Backreaction of the space--time
was considered in \cite{hhh} for (2+1)--dimensional holographic superconductor.
It was found that the qualitative results are similar for small charges, but that
suprisingly the scalar field can even form a condensate when being uncharged.

In \cite{gregory} (3+1)--dimensional superconductors were studied by 
investigating scalar hair formation on black holes in Gauss--Bonnet gravity. This has been
extended to higher dimensions in \cite{pan}.
The motivation for this is the apparent contradiction between the Mermin--Wagner theorem that forbids spontaneous symmetry
breaking in 2+1 dimensions at finite temperatures and the fact that (2+1)--dimensional
holographic superconductors do exist. Consequently, it has been suggested that
higher curvature corrections should suppress condensation, where higher curvature corretions can of course only been studied for (3+1)--dimensional superconductors (or higher dimensional ones). \cite{gregory,pan} were concerned with
the ``probe limit'' and it was found that 
condensation cannot be suppressed in Gauss--Bonnet gravity.
  
In this paper, we are interested in the model studied in \cite{gregory} but away from the probe limit, i.e. taking back--reaction of the space--time into account.
We study (3+1)-dimensional superconductors in pure Einstein and Gauss--Bonnet gravity, respectively. While for large temperatures, i.e. when the scalar field
vanishes identically analytic solutions to the equations of motion are known, this
is different for a black hole with scalar hair that forms below the condensation
temperature. These solutions have to be constructed numerically. 

In Section 2, we present the model, the equations of motion and the boundary conditions.
In Section 3, we discuss our numerical results, while Section 4 contains our conclusions.

\section{The Model}
In this paper, we are studying the formation of scalar hair on an electrically charged black hole in $(4+1)$ dimensional Anti--de Sitter space--time. 
The action reads~:
\begin{equation}
S= \frac{1}{16\pi G} \int d^5 x \sqrt{-g} \left(R -2\Lambda + 
\frac{\alpha}{4}\left(R^{\mu\nu\lambda\rho} R_{\mu\nu\lambda\rho} - 4 R^{\mu\nu} R_{\mu\nu} + R^2\right) + 16\pi G {\cal L}_{\rm matter}\right) \ ,
\end{equation}
where $\Lambda=-6/L^2$ is the cosmological constant and $\alpha$ the Gauss--Bonnet coupling.
${\cal L}_{\rm matter}$ denotes the matter Lagrangian~:
\begin{equation}
{\cal L}_{\rm matter}= -\frac{1}{4} F_{MN} F^{MN} - \left(D_M\psi\right)^* D^M \psi - m^2 \psi^*\psi  \ \ , \ \  M,N=0,1,2,3,4
\end{equation}
where $F_{MN} =\partial_M A_N - \partial_N A_M$ is the field strength tensor and
$D_M\psi=\partial_M \psi - ie A_M \psi$ is the covariant derivative.
$e$ and $m^2$ denote the electric charge and mass of the scalar field $\psi$, respectively.

The Ansatz for the metric reads~:
\begin{equation}
ds^2 = - f(r) a^2(r) dt^2 + \frac{1}{f(r)} dr^2 + \frac{r^2}{L^2} d\Sigma^2_{k,3}
\end{equation}
where $f$ and $a$ are functions of $r$ only.
The $3$-dimensional metric is
\begin{equation}
d\Sigma^2_{k,3}=
\begin{cases}
d \Omega^2_{3} \ \ \ {\rm for} \ \ k=1 \\ 
dx^2 + dy^2 + dz^2 \ \ \ {\rm for} \ \ k=0 \\ 
d \Xi^2_{3} \ \ \  {\rm for} \ \  k=-1   \ 
\end{cases}
\end{equation}
where $k$ denotes the curvature of the 3-dimensional space.
We are only interested in plane-symmetric black holes in this paper, so we will set $k=0$. However, we will keep the $k$ in the equations for completeness.

For the electromagnetic field and the scalar field we choose \cite{hhh}~:
\begin{equation}
A_{M}dx^M = \phi(r) dt \  \  \  , \   \   \   \psi=\psi(r)
\end{equation}
such that the black hole possesses only electric charge.

The coupled Einstein and Euler--Lagrange equations are obtained from the variation of the
action with respect to the matter and metric fields, respectively. They read~:
\begin{eqnarray}
\label{eq1}
     f' &=& 2r \frac{k-f+ 2 r^2/{L^2}}{r^2 + 2 \alpha(k-f)} 
     - \gamma \frac{r^3}{2 f a^2} 
     \left(\frac{2 e^2 \phi^2 \psi^2 + f (2 m^2 a^2 \psi^2 + \phi'^2) + 2 f^2 a^2 \psi'^2}{r^2 + 2 \alpha (k-f))}\right) \\
\label{eq2}
        a' &=& \gamma \frac{r^3(e^2 \phi^2 \psi^2 + a^2 f^2 \psi'^2)}{a f^2(r^2+ 2 \alpha(k-f))}\\
\label{eq3}
   \phi'' &=& - \left(\frac{3}{r} - \frac{a'}{a}\right) \phi' +2 \frac{e^2 \psi^2}{f} \phi \\
\label{eq4}
    \psi'' &=& -\left(\frac{3}{r} + \frac{f'}{f} + \frac{a'}{a}\right) \psi'- \left(\frac{e^2 \phi^2}{f^2 a^2} - \frac{m^2}{f}\right) \psi
\end{eqnarray}
where $\gamma=16\pi G$. Here and in the following the prime denotes
the derivative with respect to $r$. In \cite{gregory}, $(3+1)$-dimensional holographic superconductors have been studied in the probe limit corresponding to $\gamma=0$. For $\gamma\neq 0$ we take back--reactions
of the space--time into account.
Note that this limit is equivalent to letting $e\rightarrow\infty$ since
we can preform the rescalings $\psi \rightarrow \psi/e$, $\phi\rightarrow \phi/e$
and $\gamma \rightarrow e^2\gamma$. Hence without loosing generality we can set $e\equiv 1$. 

In order to find an explicit solution of the equations of motion, we have
to fix appropriate boundary conditions. 
In the following, we are interested in the formation of scalar hair on
electrically charged black holes with horizon at $r=r_h$ such that
\begin{equation}
\label{bc1}
 f(r_h)=0  
\end{equation}
with $a(r_h)$ finite. 
In order for the matter fields to be regular at the horizon we need to impose:
\begin{equation}
\label{bc2}
\phi(r_h)=0 \ \ , \ \ \psi'(r_h) = \left.\frac{ m^2 \psi\left(r^2+2\alpha k\right)}  {2rk + 4r/L^2 -\gamma r^3 \left( m^2 \psi^2 +  \phi'^2/(2a^2)\right)   }\right\vert_{r=r_h}   \ .
\end{equation}
Asymptotically, we want the space--time to be that of a 
Reissner-Nordstr\"om--Anti de Sitter black hole, i.e. we can choose
$a(r\rightarrow\infty)\rightarrow 1$. Other choices of the asymptotic value
of $a(r)$ would simply correspond to a rescaling of the time coordinate. The matter fields on the other hand obey \cite{gregory}~:
\begin{equation} 
  \phi(r\gg 1) = \mu - \rho/r^2  \ \ , \ \ 
  \psi(r\gg 1) = \frac{\psi_{-}}{r^{\lambda_{-}}} + \frac{\psi_{+}}{r^{\lambda_{+}}} \ \
\label{decay}
\end{equation}
with
\begin{equation}
\label{lambda}
       \lambda_{-} = 2 - \sqrt{4- 3 (L_{\rm eff}/L)^2} \ \ , \ \ \lambda_{+} = 2 + \sqrt{4- 3 (L_{\rm eff}/L)^2} \ \ , \ \ 
       L_{\rm eff}^2 \equiv \frac{2 \alpha}{1 - \sqrt{1 - 4 \alpha/L^2}} 
       \sim L^2 \left(1  -  \alpha/ L^2 + O(\alpha^2)\right)   \ .
\end{equation}
Note that the value of the Gauss--Bonnet coupling $\alpha$ 
is bounded from above~: $\alpha \leq L^2/4$ where $\alpha=L^2/4$ is the Chern-Simons limit. For larger values of
$\alpha$ the solution would possess a naked singularity. 

The parameters $\mu$, $\rho$ are the chemical potential and density of electric charge, respectively. Along with \cite{gregory} we choose $\psi_{-}=0$.
$\psi_{+}$ will correspond to the expectation value $\langle{\cal O}\rangle$ of the operator
${\cal O}$ which in the context of the gauge theory--gravity duality is dual
to the scalar field and as such represents the value of the condensate. 

There are analytic solutions of the equations of motion for $\psi(r)\equiv 0$~:
\begin{equation}
      f(r) = k + \frac{r^2}{2\alpha} \left(1-\sqrt{1-\frac{4\alpha}{L^2} + \frac{4\alpha M}{r^4} - \frac{4\alpha\gamma\rho^2}{r^6}}\right)  \ \ , \ \ a(r)=1 \ \ , \ \ 
      \phi(r) = \frac{\rho}{r_h^2} - \frac{\rho}{r^2}
\label{rn}
\end{equation}
where $M$ and $\rho$ are arbitrary integrations constants that can be interpreted as the mass and
the charge density of the solution, respectively. In the limit $\alpha\rightarrow 0$, the metric function $f(r)$ becomes $f(r)=k +\frac{r^2}{L^2}-\frac{M}{r^2}+\frac{\gamma\rho^2}{r^4}$.
These solutions are electrically charged black holes which are the only solutions for temperatures larger
than the critical temperature $T_c$. For $T < T_c$ these solutions
will be unstable to form scalar hair, i.e. develop a non-vanishing
value of $\psi$ on the horizon. The aim of this paper is to study the formation
of scalar hair black holes in dependence on $\alpha$ and $\gamma$.
The temperature mentioned here corresponds to the Hawking temperature of the
black hole and reads
\begin{equation}
          T = \left.\frac{1}{4\pi}\sqrt{-g^{tt} g^{MN} \partial_M g_{tt} \partial_N g_{tt}}\right\vert_{r=r_h}= \frac{1}{4 \pi} f'(r_h) a(r_h)  \ , \ \ M,N=1,2,3,4  \ .
\end{equation}
In the gauge theory -- gravity duality $T_c$ is the temperature below which
superconductivity appears.

\section{Numerical results}
\label{numerics}

In the following we are only interested in the plane--symmetric black holes
with $k=0$. 
The equations of motion (\ref{eq1})-(\ref{eq4}) depend in principle on a number of constants but due to the scale invariances noted in \cite{hhh}
two of them can be scaled out and hence be fixed to particular values without
loosing generality. In the following we fix $r_h=0.5$ and $L=1$.
Along with \cite{gregory} we set $m^2=-3/L^2\equiv -3$  which guarantees the
stability of $AdS_5$ since $m^2 < m^2_{\rm BF}= -4/L^2$ with $m^2_{\rm BF}$ the Breitenlohner--Freedam mass \cite{bf}. 

To find a unique solution to the equations
of motions we fix the boundary conditions (\ref{bc1}), (\ref{bc2}) at the horizon, and choose $a(\infty)=1$, $\psi_{-}=0$. In addition we fix $\psi_{+}$ to a particular value. In this way we are able to construct branches of solutions
labelled by the parameter $\psi_{+}$. 
Note that $\mu$ and $\rho$ are uniquely fixed by the choice of
$\psi_{+}$ and are not free parameters. 
However, in the literature on holographic superconductors, the solutions
are typically presented for fixed electric charge density $\rho$, while the horizon value $r_h$ is treated as a free parameter. These two approaches are connected
to each other. Indeed, it is
easy to convert a branch of solutions with fixed $r_h$ and varying $\psi_{+}$ -- where
$\rho=\rho(\psi_{+})$ -- into a branch of solutions with constant
charge density. In the following we denote quantities corresponding to a fixed charge
density by a hat. Setting $\hat\rho=1$, 
the relevant Hawking temperature $\hat T$ and condensate $\hat \psi_{+}$ are respectively given by
\begin{equation}
\label{physic_scaling}
     \hat T = \frac{T}{\rho^{1/3}} \ \  , \ \ \hat \psi_{+} = \frac{\psi_{+}}{\rho^{(\lambda_{+}/3)}} \ \ , 
\end{equation}  
where $\lambda_{+}$ is defined in (\ref{lambda}).


\subsection{Effect of back--reaction in Einstein gravity}
This corresponds to the case $\gamma\neq 0$ and $\alpha=0$.
We solved the equations for several values of $\gamma$ and $\psi_{+}$ and
find that solutions exist for generic values of these parameters. 

When studying solutions for $\gamma$ fixed and varying
$\psi_{+}$ we find that in the limit $\psi_{+} \to 0$ the 
solutions tend to (\ref{rn}) for very specific values of $\mu$ and $\rho$ which
depend on the choice of $\gamma$ and can only be determined numerically.
Correspondingly, the critical temperature $\hat{T}_c$ at which $\hat{\psi}_{+}=0$ can also only be determined numerically. 
We find the values given in the table below~:
\begin{center}
\begin{tabular}{|c||c|c|c|c|c|c|c|c|}
\hline
$\gamma$ & $0.0$ & $0.025$ & $0.05$ & $0.1$ & $0.15$ & $0.2$ & $0.3$ & $0.35$ \\
\hline
$4\pi \hat{T}_c$ & $2.48 $ & $2.02 $  & $1.61 $ & $0.99 $ & $0.57$ & $0.33 $ & $0.10 $ & $0.06$  \\
\hline
\end{tabular}
\end{center}

For large $\gamma$ the construction of the solutions becomes increasingly
difficult. In principle we would want to know what happens for very large
$\gamma$. In order to understand this, we fitted the numerical data and found that
\begin{equation}
 T_c \approx 0.198 \cdot \exp\left(-10.6 \cdot \gamma\right)\rho^{1/3}
\end{equation}
fits the data for $\gamma \ge 0.2$ very well.
This on the other hand means that no matter how large we choose $\gamma$, we will always have $T_c > 0$.  This
has already been observed for superconductors in $(2+1)$ dimensions \cite{hhh}, where it was shown
that $T_c > 0$ in the limit $e\rightarrow 0$ which corresponds to $\gamma\rightarrow\infty$ here.
Apparently, this phenomenon persists for $(3+1)$-dimensional superconductors.


Fixing $\gamma$ and increasing  the value of the condensate $\psi_{+}$ we find that the values $a(r_h)$ and $\phi'(r_h)$
slowly approach zero. At the same time, the function $f(r)$ develops a local maximum and
a local minimum at values $r_{\rm M}$, $r_{\rm m}$ such that $r_h < r_{\rm M} < r_{\rm m} < \infty$.
This is illustrated for the metric functions $f(r)$ and $a(r)$ in Fig. \ref{fig4} (left) for $\gamma=0.2$ and three different values of $\psi_{+}$.

\begin{figure}[ht]
\hbox to\linewidth{\hss%
	\resizebox{9cm}{7cm}{\includegraphics{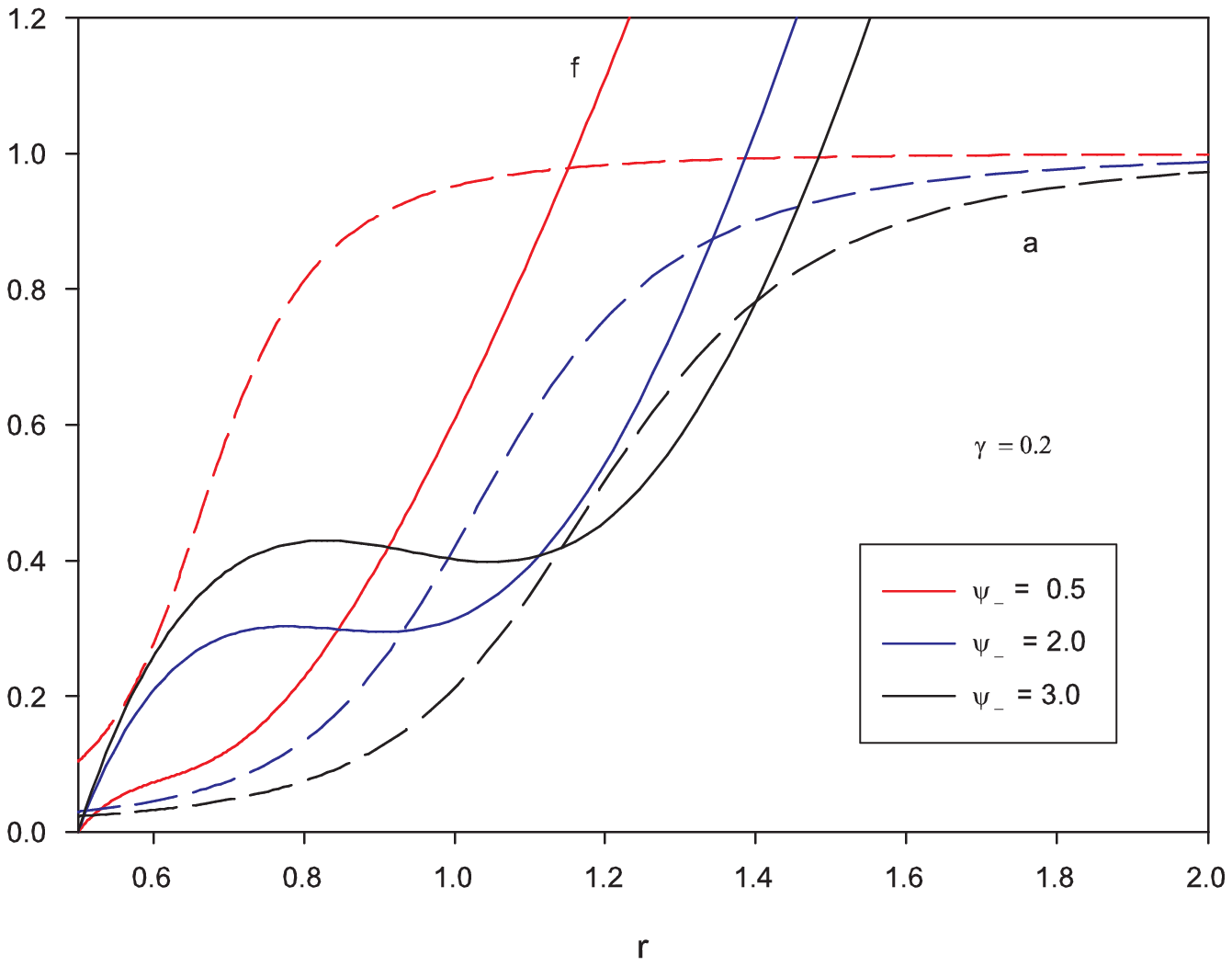}}
\hspace{5mm}%
        \resizebox{9cm}{7cm}{\includegraphics{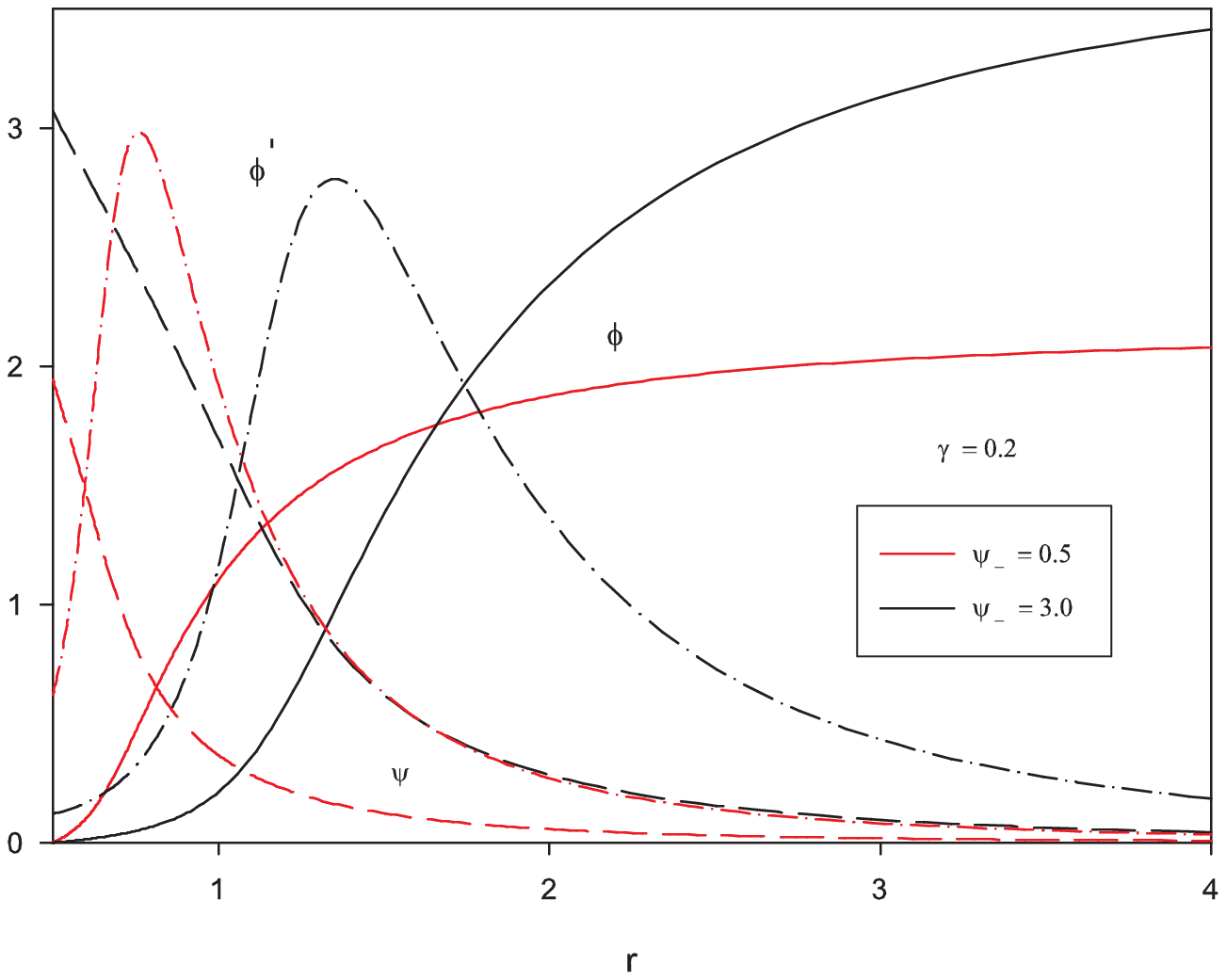}}	
\hss}
 
\caption{
{\small
(left) The metric functions $f(r)$  and $a(r)$ (left) and the matter
functions $\phi(r)$, $\psi(r)$ and $\phi'(r)$ (right) for $\gamma = 0.2$ and for three values of $\psi_{+}$.
}
 }
\label{fig4}
\end{figure}
\begin{figure}[ht]
\hbox to\linewidth{\hss%
	\resizebox{10cm}{8cm}{\includegraphics{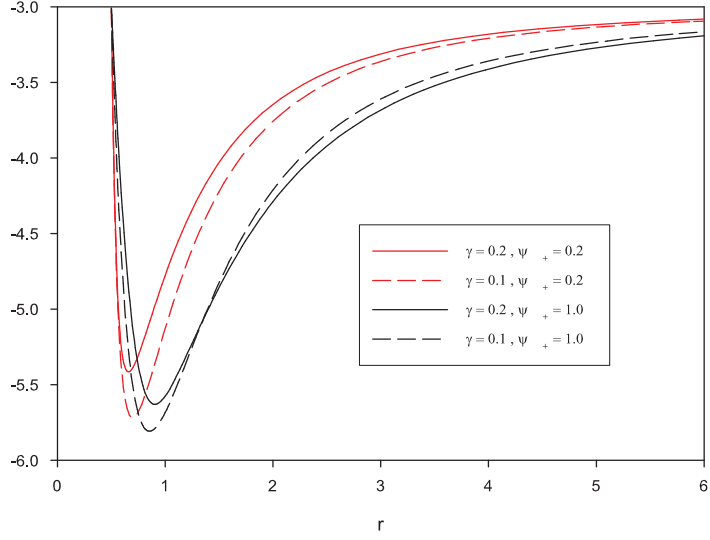}}
\hss}
 
\caption{
{\small
The value of the effective mass $m_{\rm eff}^2 = m^2 + e^2 A_t^2 g^{tt} = - \frac{3}{L^2} - \frac{\phi^2}{fa^2}$ close to the horizon of the black hole for several
values of $\gamma$ and $\psi_+$.}}
\label{fignew}
\end{figure}
\begin{figure}[ht]
\hbox to\linewidth{\hss%
	\resizebox{9cm}{7cm}{\includegraphics{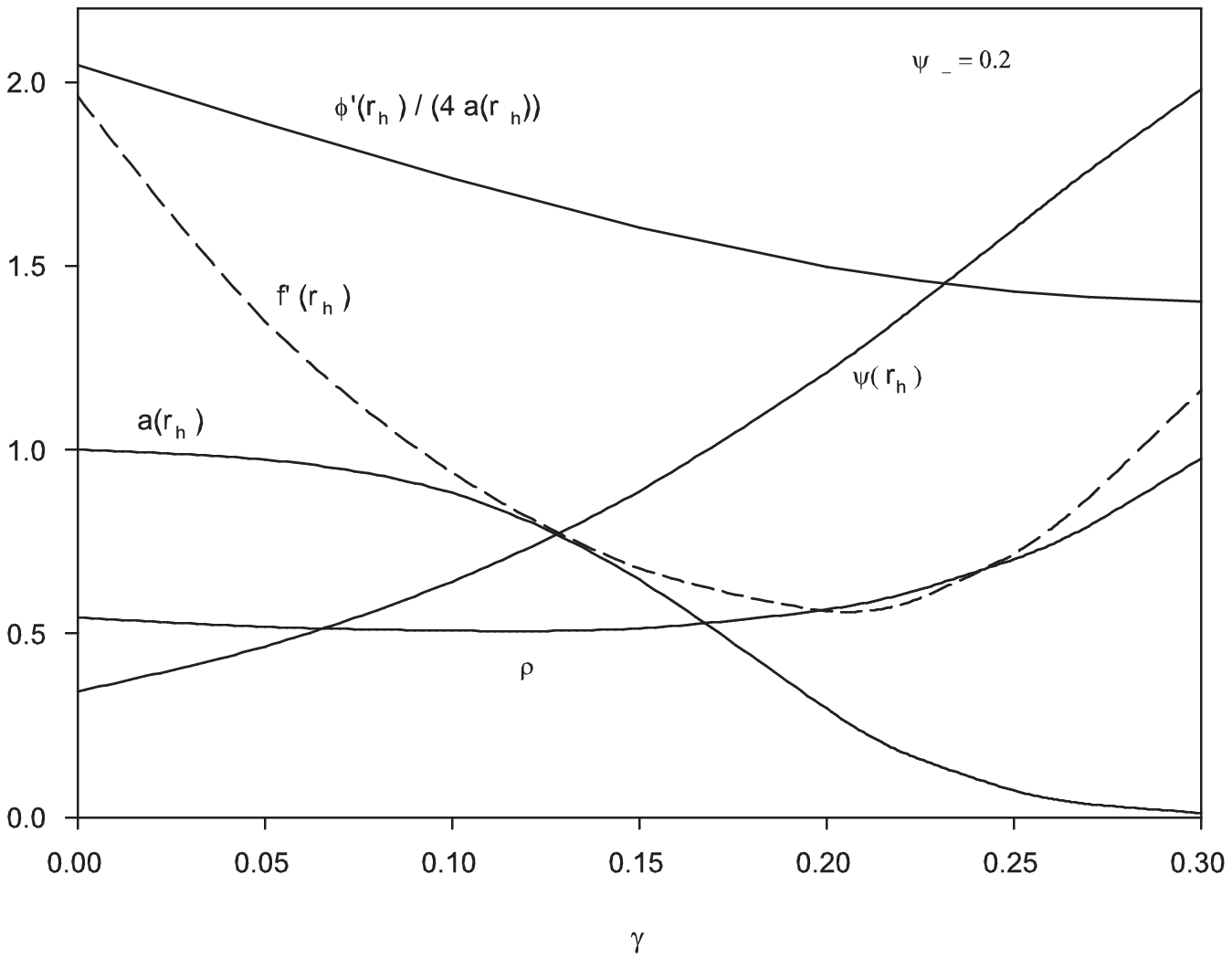}}
\hspace{5mm}%
        \resizebox{9cm}{7cm}{\includegraphics{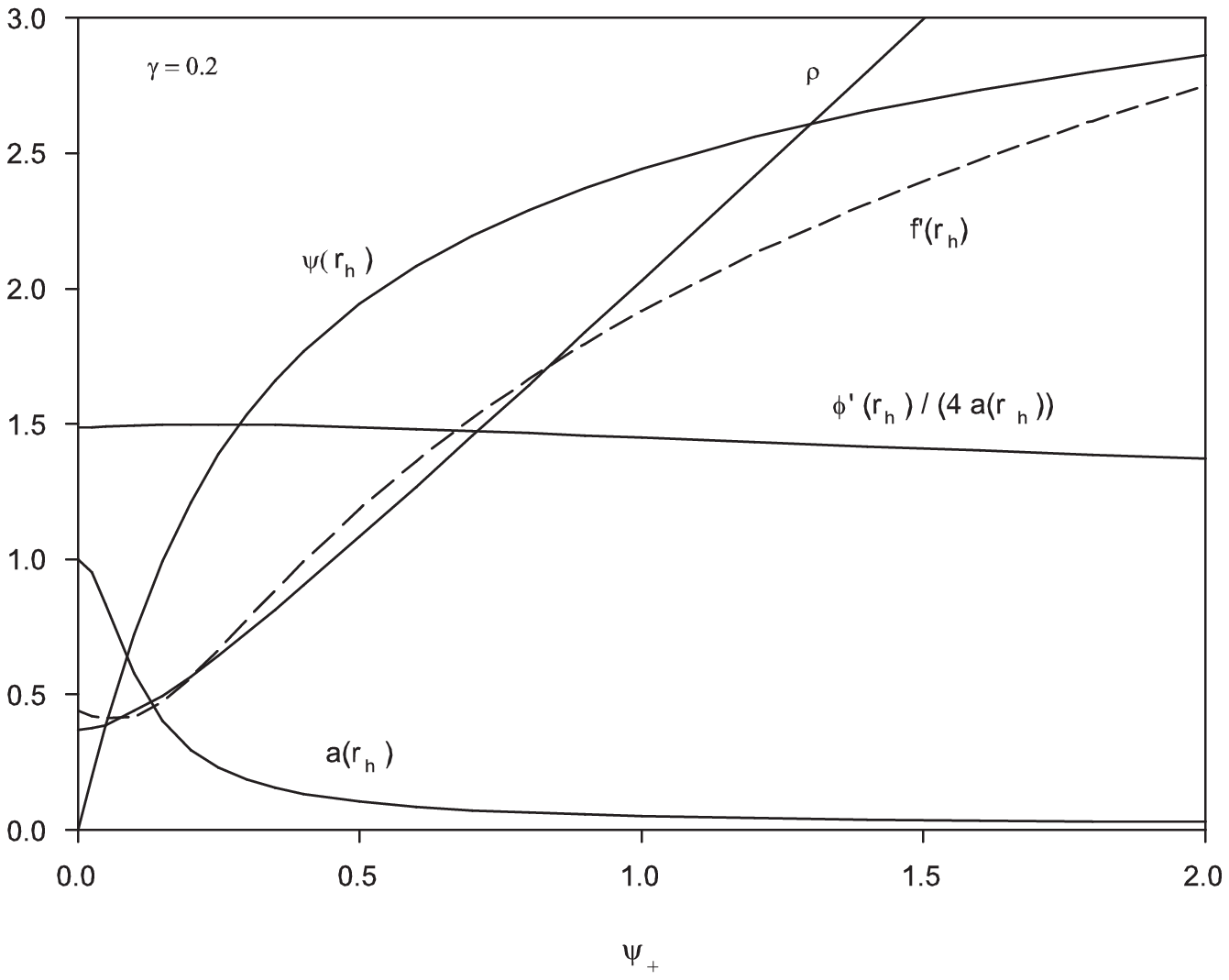}}	
\hss}
 
\caption{
{\small
Several quantities characterizing the black holes in pure Einstein gravity ($\alpha=0$) with
$\psi_+= 0.2$ for $\gamma$ varying (left) and for $\gamma=0.2$ and $\psi_+$ varying (right).
}
 }
\label{fig3}
\end{figure}
\begin{figure}[ht]
\hbox to\linewidth{\hss%
	\resizebox{9cm}{7cm}{\includegraphics{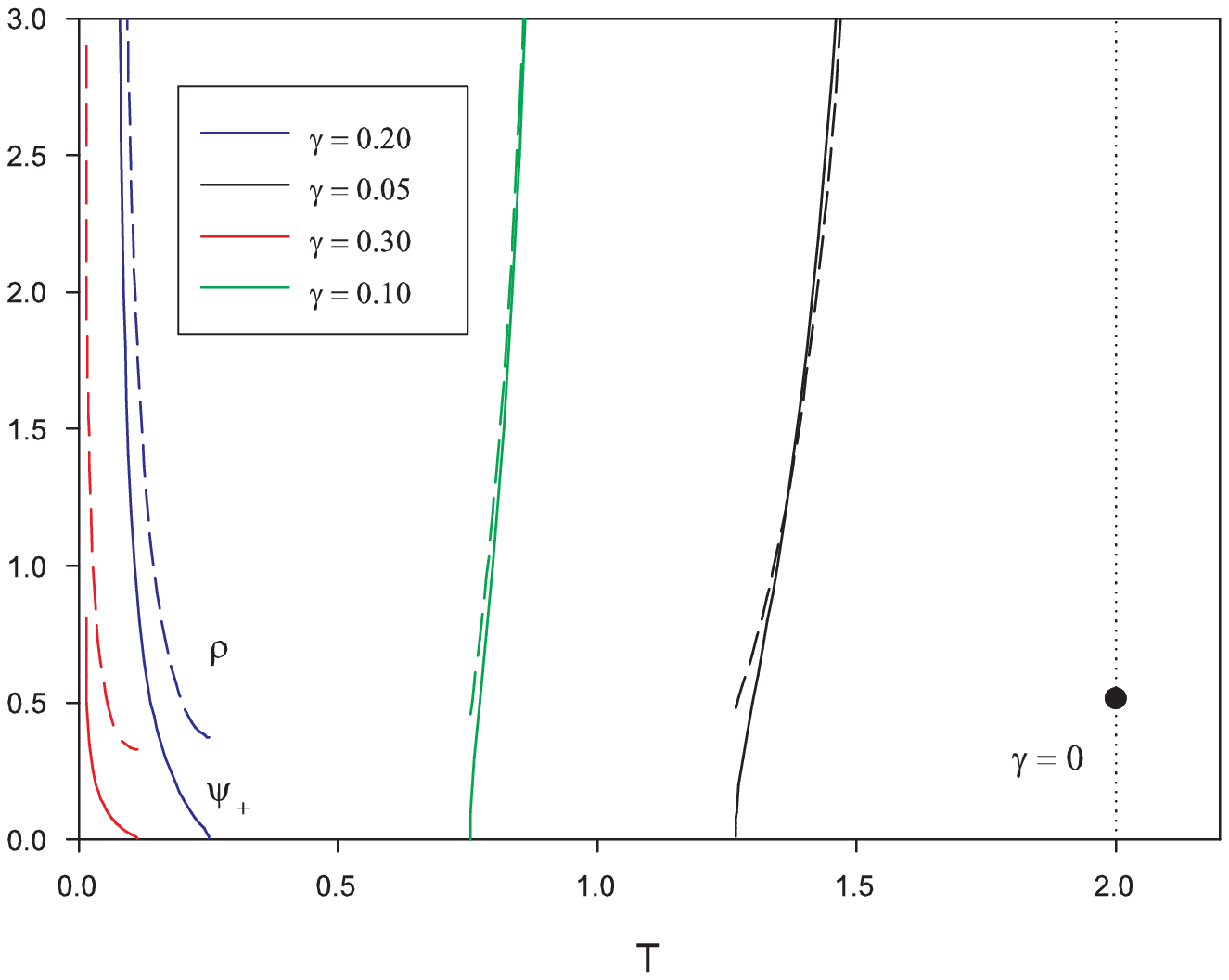}}
\hspace{5mm}%
        \resizebox{9cm}{7cm}{\includegraphics{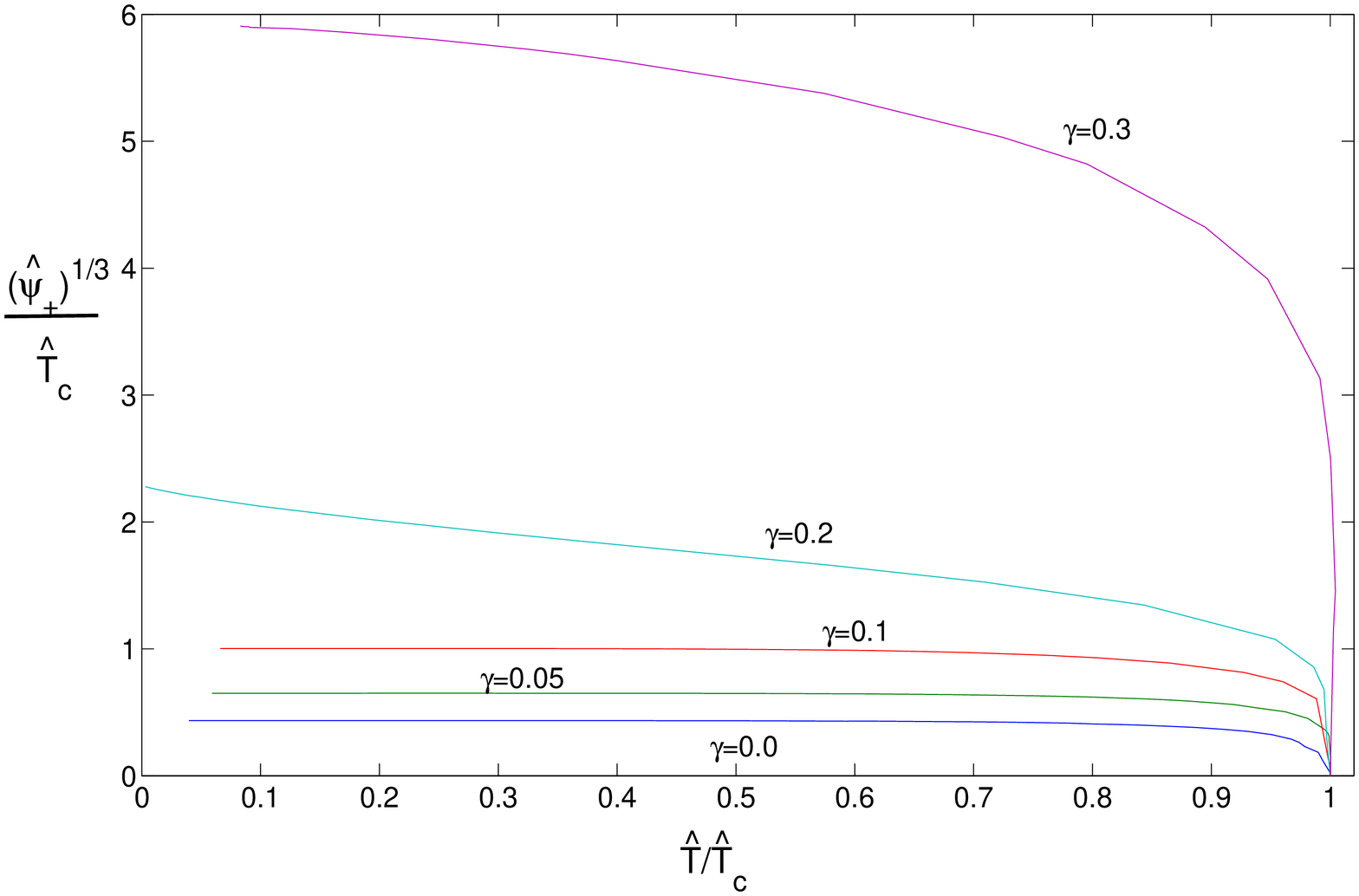}}	
\hss}
 
\caption{
{\small
The values $\psi_+$ and $\rho$ as functions of the temperature $T$ for several values of $\gamma$ with $r_h=0.5$ (left). 
The values $(\hat{\psi}_+)^{1/3}/\hat{T}_c$  as function of $\hat{T}/\hat{T}_c$ for several values of $\gamma$ with $\hat{\rho}=1$ (right).
}
 }
\label{fig_tc_rh_05}
\end{figure}
\begin{figure}[ht]
\hbox to\linewidth{\hss%
	\resizebox{10cm}{8cm}{\includegraphics{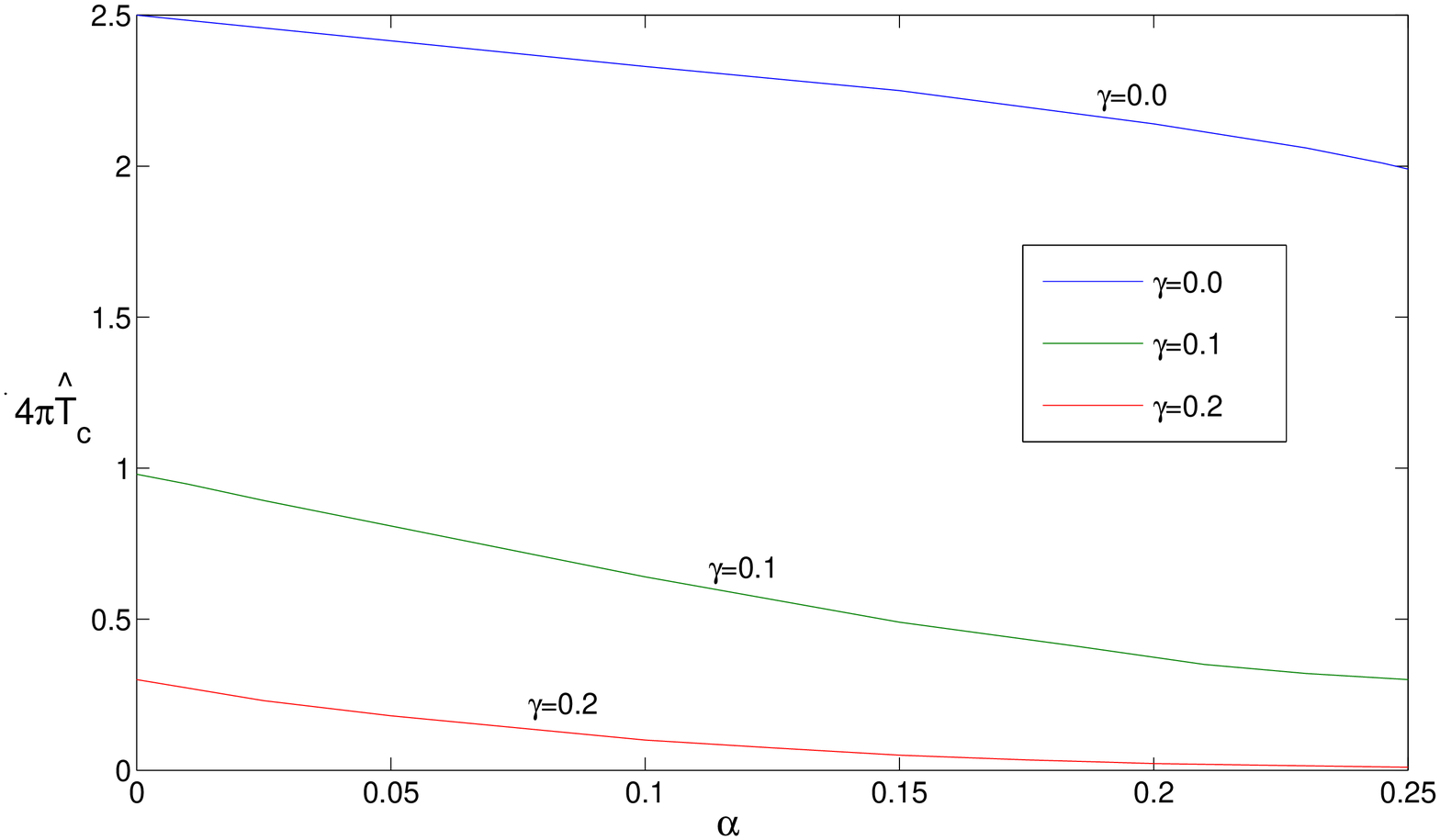}}
\hss}
 
\caption{
{\small
The critical temperature $\hat{T}_c$ at which superconductivity sets in as function of the Gauss--Bonnet coupling constant $\alpha$ for several values of the gravitational
coupling $\gamma$.
}
 }
\label{critical}
\end{figure}
 

This would suggest that for sufficiently large $\psi_{+}$ the function $f(r)$ develops a double zero at $r=r_{\rm m}$
which would correspond to the formation of an extremal black hole with vanishing Hawking temperature.
A detailed analysis however shows that the value
$f(r_m)$ remains strictly positive,  while 
the value of  $a(r_h)$ decreases with $\psi_{+}$ increasing according to an exponential behaviour
$a(r_h)\sim \exp\left(-c\psi_{+}\right)$ with $c$ some constant.
This result  suggests that the black hole solutions are not limited by a maximal value of the condensate $\psi_{+}$ and that the temperature stays positive for all values of
$\psi_+$.

In Fig.\ref{fig4} (right) we show the matter field functions $\phi(r)$, $\psi(r)$ and $\phi'(r)$. We observe that 
for fixed $\gamma$ and increasing $\psi_{+}$  the maximum of $\phi'(r)$ is pushed further away from
the horizon of the black hole. Indeed, for small values of $\gamma$ and $\psi_{+}$, the maximum of $\phi'(r)$ is on the horizon of the black hole. 
Note that when fixing $\psi_{+}$  and increasing $\gamma$ we observe a similar phenomenon.

To understand how the scalar field $\psi$ becomes unstable
close to the horizon, we plot the effective mass
\begin{equation}
m_{\rm eff}^2 = m^2 + e^2 A_t^2 g^{tt} = - \frac{3}{L^2} - \frac{\phi^2}{fa^2} 
\end{equation}
in Fig.\ref{fignew} for $L=1$. 
Indeed, the effective mass drops below the Breitenlohner-Freedman bound of $-4/L^2=-4$ 
close to the horizon. For fixed $\psi_+$ and increasing $\gamma$ the quantity $- \frac{3}{L^2} - \frac{\phi^2}{fa^2}$
becomes more narrow and smaller in absolute value. For fixed $\gamma$ and increasing $\psi_+$
it becomes broader and larger in absolute value.

To understand this in more detail note that close to the horizon the functions can be expanded as follows~:
\begin{eqnarray}
 f(r) &=& f'(r_h)(r-r_h) + \frac{f''(r_h)}{2}(r-r_h)^2 + ... \\
a(r) &=& a(r_h) + a'(r_h)(r-r_h) + \frac{a''(r_h)}{2}(r-r_h)^2 + ... \\
\phi(r) &=& \phi'(r_h)(r-r_h) + \frac{\phi''(r_h)}{2}(r-r_h)^2 + ...\\
\psi(r) &=& \psi(r_h) + \psi'(r_h)(r-r_h) + \frac{\psi''(r_h)}{2}(r-r_h)^2 + ...  \ .
\end{eqnarray}
Moreover note that there are the following relations between the values of the
functions at $r=r_h$:
\begin{equation}
\label{phirh}
\phi''(r_h)=\phi'(r_h)\left(\frac{a'(r_h)}{a(r_h)} + 2e^2 \frac{\psi(r_h)^2}{f'(r_h)}-\frac{3}{r_h} \right)
\end{equation}
\begin{equation}
\label{psirh}
\psi'(r_h) f'(r_h) = m^2 \psi(r_h)=-\frac{3}{L^2} \psi(r_h) \
\end{equation}
\begin{equation}
\label{frh}
f'(r_h)= 4 \frac{r_h}{L^2}-r_h \gamma\left(  m^2 \psi(r_h)^2 + \frac{\phi'(r_h)^2}{2a(r_h)^2}\right)= 
4 \frac{r_h}{L^2}+r_h \gamma\left( \frac{3}{L^2} \psi(r_h)^2 - \frac{\phi'(r_h)^2}{2a(r_h)^2}\right)
 \ ,
\end{equation}

\begin{equation}
\label{arh}
a'(r_h)=r_h\gamma \left(a(r_h) \psi'(r_h)^2+e^2 \frac{\phi'(r_h)^2 \psi(r_h)^2}{a(r_h)f'(r_h)^2}\right)
\end{equation}

Several quantities characterizing the solutions
are given in Fig.\ref{fig3}  for  $\gamma$ varying
and $\psi_+=0.2$ fixed (left) and for $\psi_+$ varying  and $\gamma=0.2$ fixed (right), respectively. First note that $\psi(r_h)$ increases with $\gamma$ and $\psi_+$, respectively.
That $\psi(r_h)$ is an increasing function of the condensate $\psi_+$ 
was already noticed in \cite{gregory} for the probe limit. Here, we find in addition that 
the stronger the back--reaction the higher the value $\psi(r_h)$
for a given condensate $\psi_+$. Since we would like $\psi$ to have its maximal positive value on the horizon we have $\psi(r_h)>0$,  $\psi'(r_h) < 0$ and then from (\ref{psirh}) obviously $f'(r_h) > 0$. For $\gamma=0$, the value of $f'(r_h)=4 \frac{r_h}{L^2}$ which for $r_h=0.5$ and $L=1$ is just $f'(r_h)=2.0$. For increasing $\gamma$ the value $f'(r_h)$ is first decreasing
due to the decrease of the electric field $\phi'(r_h)/a(r_h)$ on the horizon (the negative
term in (\ref{frh})) and then for sufficiently strong back--reaction $\psi(r_h)$ becomes larger and larger such that $f'(r_h)$ starts increasing again.
This is similar for $\gamma$ fixed and varying $\psi_+$. For $\psi_+=0$, the solution
is given by (\ref{rn}). For $\psi_+ > 0$ but small the electric field on the horizon
$\phi'(r_h)/a(r_h)$ decreases and leads to a slight decrease in $f'(r_h)$. For increasing $\psi_+$ the value $\psi(r_h)$ becomes larger and $f'(r_h)$ increases.
For both $\gamma$ and $\psi_+$ increasing, respectively, the value of $a(r_h)$ decreases
from $a(r_h)=1$. 
 
Finally, the charge density $\rho$ decreases for small $\psi_+$ 
and increases for larger $\psi_+$ when $\gamma$ is fixed. For fixed $\psi_+$ and varying
$\gamma$ the behaviour is qualitatively similar. 

When studying the dependence of the condensate $\psi_+$ on the temperature $T$ one can take two different
viewpoints. First we consider the system for fixed horizon ($r_h= 0.5$ here).
The dependence of $<{\cal O}> = \psi_+$ as a function of the temperature $T$ is given in 
Fig. \ref{fig_tc_rh_05} (left) for several values of $\gamma$. 

The solid line represents the condensate $\psi_+$ and the dashed line
represents the charge density $\rho$, respectively. For completeness we give the corresponding lines for $\gamma=0$ (dashed line), where the bullet
indicates the minimal value of $\rho$. Apparently, the behaviour is quite different when comparing 
large and small $\gamma$.
For small $\gamma$ the temperature of the condensate is larger than the temperature of the critical limit,
for large $\gamma$ it is vice versa.

Following the literature, we also present our results for fixed charge density $\hat{\rho}=1$.
The dimensionless quantity $<{\cal \hat{O}}>^{1/\lambda_+}/\hat{T}_c=(\hat{\psi}_+)^{1/\lambda_+}/\hat{T}_c$
(with $\lambda_+=3$ in the limit $\alpha=0$) as a function of 
the rescaled temperature $\hat{T}/\hat{T}_c$ is given in Fig. \ref{fig_tc_rh_05} (right).
Qualitatively, the behaviour for large $\gamma$ is similar to that for small $\gamma$.
However, the condensate can become quite large when increasing $\gamma$.
Moreover, the critical temperature $\hat{T}_c$ at which $\hat{\psi}_+=0$  decreases with increasing $\gamma$.

Note that we are in fact plotting $(e\hat{\psi}_+)^{1/\lambda_+}/\hat{T}_c$ but that
$e$ doesn't appear here due to our choice $e\equiv 1$. Moreover, comparing
our results to the (2+1)-dimensional case \cite{hhh} our choices
of $\gamma=0.0$, $0.05$, $0.1$, $0.2$ and $0.3$, respectively would
correspond to $e=\infty$, $4.47$, $3.16$, $2.24$ and $1.83$ when setting $\gamma=1$
instead of $e=1$.

\subsection{Effect of back--reaction in Gauss--Bonnet gravity}
This corresponds to the case $\gamma\neq 0$ and $0 < \alpha \le 0.25$.
The $\gamma=0$ limit was studied in \cite{gregory}. Our numerical results indicate that
also for $\gamma\neq 0$ the presence of the Gauss--Bonnet term leads to a decrease
in the critical temperature.
This is shown in Fig. \ref{critical} where we give the dependence of $\hat{T}_c$
on $\alpha$ for different values of $\gamma$. 
Apparently also back--reaction on the space--time cannot suppress condensation, i.e. the critical temperature stays positive for all values of $\gamma$ and $\alpha$ that we have studied in this paper. We find e.g. for $\alpha=0.1$ that $T_c=0.185\rho^{1/3}$ for
$\gamma=0$ (in agreement with \cite{gregory}), $T_c=0.051\rho^{1/3}$ for $\gamma=0.1$
and $T_c=0.008\rho^{1/3}$ for $\gamma=0.2$. For $\alpha=0.25$ we find
$T_c=0.158\rho^{1/3}$ for
$\gamma=0$ (again in agreement with \cite{gregory}), $T_c=0.024\rho^{1/3}$ for $\gamma=0.1$
and $T_c=0.001\rho^{1/3}$ for $\gamma=0.2$. Hence, the critical temperature can become
arbitrarily close to zero for $\alpha$ and $\gamma$ large enough, however within our numerical accuracy, we never find $T_c=0$ for finite values of $\gamma$ and $\alpha \le 0.25$.

\section{Conclusions}
In this paper, we have studied holographic superconductors in 3+1 dimensions
away from the probe limit. We considered the case of pure Einstein and Gauss--Bonnet
gravity, respectively and have constructed numerically electrically charged black holes
that carry scalar hair. For pure Einstein gravity we find that in agreement with the results
for holographic superconductors in 2+1 dimensions \cite{hhh}
the critical temperature at which condensation sets in is strictly positive
for all values of the gravitational coupling. Considering Gauss--Bonnet corrections
further decreases the critical temperature, but all our numerical results indicate
that it stays positive when taking back--reaction into account. Hence, even when taking
the gravitational coupling to infinity -- which corresponds to letting the electric charge
$e$ of the condensate tend to zero -- there would still be condensation.
Similar to 2+1 dimensions this signals the existence of an additional
instability of the scalar field. The explanation is similar to that in 2+1 dimensions \cite{hhh,reviews}: since the scalar field is uncharged the instability cannot be caused
by the spontaneous symmetry breaking. Rather it is caused by the fact that
for large $\gamma$ the black hole is close to the extremal limit in which
its horizon geometry would correspond to $AdS_2\times \mathbb{R}_3$. In $AdS_2$ the Breitenlohner-Freedman bound \cite{bf} is $m^2_{BF}=-1/(4L^2)$. Hence a scalar field with
mass $m^2=-3/L^2$ that is stable in $AdS_5$ is certainly unstable in $AdS_2$.

When considering non--abelian holographic superconductors it has been
observed that the phase transition that leads to the formation of vector hair
becomes first order if the gravitational coupling is large enough \cite{aegko}. It would be
interesting to see how the Gauss--Bonnet term influences this result.
\\
\\
{\bf Acknowledgments} We thank
J. Erdmenger for comments on this manuscript. YB thanks the Belgian FNRS for financial support.

\end{document}